\documentclass[draft]{agujournal2019}
\usepackage{url} 
\usepackage{lineno}
\usepackage[inline]{trackchanges} 
\usepackage{soul}

\draftfalse

\journalname{Geophysical Research Letters}

\begin{document}

%
%


\title{Fe\textsuperscript{2+} partitioning in Al-free pyrolite: consequences for seismic velocities and heterogeneities}

%
%

\authors{Jingyi Zhuang\affil{1,2}, Renata Wentzcovitch\affil{1-5}}

\affiliation{1}{Department of Earth and Environmental Sciences, Columbia University, New York, NY 10027, US}
\affiliation{2}{Lamont–Doherty Earth Observatory, Columbia University, Palisades, NY 10964, US}
\affiliation{3}{Department of Applied Physics and Applied Mathematics, Columbia University, New York, NY 10027, US}
\affiliation{4}{Data Science Institute, Columbia University, New York, NY 10027, US}
\affiliation{5}{Center for Computational Quantum Physics, Flatiron Institute, New York, NY 10010, US}


\correspondingauthor{Renata Wentzcovitch}{rmw2150@columbia.edu}



\begin{keypoints}
\item The ferrous iron spin crossover (ISC) affects the iron partitioning between the lower mantle phases, bridgmanite and ferropericlase.

\item Iron partitioning effects on mantle velocities are small but manifest clearly on thermally induced velocity heterogeneity ratios.

\item The predicted S- to P-velocity heterogeneity ratio is similar to those inferred from tomography models down to 2,400 km depth.

\end{keypoints}

\begin{abstract}

Iron partitioning among the main lower mantle phases, bridgmanite (Bm) and ferropericlase (Fp), has non-monotonic behavior owing to the high-spin to low-spin crossover in ferrous iron (Fe\textsuperscript{2+}) in Fp. Results of previous studies of the iron partitioning coefficient between these phases, $K_D$, still have considerable uncertainty. Here, we investigate the Fe\textsuperscript{2+} partitioning behavior using well-documented \textit{ab initio} free energy results plus new updates. Although we focus on Fe\textsuperscript{2+} only, we describe the effect of this iron spin crossover (ISC) on $K_D$ and of the latter on compositions and seismic velocities in a pyrolitic aggregate. Our results suggest that its velocities are mainly affected by the ISC and less so by the Fe\textsuperscript{2+} partitioning. In contrast, iron partitioning manifests in thermally induced velocity heterogeneity ratios. Prediction of the seismological parameter $R_{S/P}$ ($\partial \ln V_S/\partial \ln V_P$) including iron partitioning effects resembles quantitatively $R_{S/P}$'s inferred from several tomographic studies down to 2,400 km depth.

\end{abstract}

\section*{Plain Language Summary}
Since the discovery of the iron spin crossover in lower mantle phases, its effect on mantle properties has been a matter of debate. The ferrous iron partitioning coefficient between the main lower mantle phases, bridgmanite (Bm) and ferropericlase (Fp), is the starting point for understanding these effects. Here, we show that iron partitioning changes due to the ISC in Fp do not affect significantly the velocities of a typical pyrolitic aggregate. However, it can cause a perceptible effect on the thermally induced S- to P-velocity heterogeneity ratio, i.e., an undulated increase with depth similar to what is inferred from some tomography models down to ~2,400 km depth.

%
%
\clearpage
\section{Introduction}

The lower mantle consists primarily of the iron-bearing phases bridgmanite (Bm) ((Mg, Al, Fe)(Si,Al,Fe)O\textsubscript{3}-perovskite), its polymorph post-perovskite (PPv), and ferropericlase (Fp) (B1-type (Mg,Fe)O). 
Cubic CaSiO\textsubscript{3}-perovskite (cCaPv) should also be present as an independent phase in the upper lower mantle, but it has been shown to dissolve into Bm in the mid to deep lower mantle \cite{Muir2021TheRegions, Ko2022CalciumMantle}. 
The iron partitioning coefficient, $K_D$, between the silicate and the oxide phases is a key parameter that affects phase compositions, aggregate properties, and, possibly and indirectly, the mantle dynamic behavior \cite{Dorfman2021CompositionHeterogeneities}. 
The discovery of the iron spin crossover (ISC) from high-spin (HS) to low-spin (LS) in these phases \cite{Badro2003, Badro2004ElectronicMantle} immediately raised questions about its consequences for mantle properties, starting with its effect on $K_D$. 
The theoretical description of the ISC \cite{Tsuchiya06} and related elastic anomalies \cite{Wentzcovitch2009, Wu2013}, elevate the complexity of thermochemical equilibrium and seismic wave velocity calculations in lower mantle aggregates, already very challenging problems.

This paper addresses the effect of the ISC in ferrous iron (Fe\textsuperscript{2+}) in Fp on the Fe\textsuperscript{2+} partitioning in an aggregate with a typical pyrolitic Mg/Si ratio, $\sim 1.27$, and average iron concentration $x^\mathrm{Fe} = 0.1$. 
While no ferric iron (Fe\textsuperscript{3+}), Al or Ca are included in this analysis and, therefore, it does not address the full complexity of the mantle, this more limited analysis allows for great control in identifying causes and consequences of the Fe\textsuperscript{2+} partitioning on seismic velocities. 
The latter has been an urgent question relevant to the analyses of tomography models attempting to identify the signature of the ISC \cite{Shephard2021SeismologicalMantle, Trautner2023CompressibilityTomography, Cobden2024Full-waveformMantle}. 

Experimental measurements of $K_D$ on complex aggregates with likely mantle compositions offer a range of values with considerable uncertainties. 
Essentially, two classes of aggregates have been extensively investigated: aggregates containing Al \cite{Irifune2, Irifune2010, Kesson1998, Murakami2005, Sinmyo2013, Prescher2014, Wood2000, piet} 
and Al-free systems \cite{sakai09, sinmyo08, auzende, kobayashi05, piet}. 
The latter is assumed to contain only Fe\textsuperscript{2+} substituting for Mg ($\sim$(Mg\textsubscript{0.9},Fe\textsubscript{0.1})SiO\textsubscript{4}). 
In short, $K_D$ is larger in Al-bearing aggregates, $\sim 0.5$, since Al increases the Fe\textsuperscript{3+} content in the system \cite{Irifune2010}. 
In Al-free systems, $K_D$ is smaller ($\sim 0.1- 0.3$) (see, e.g., \cite{piet}). 
Aiming at shedding light on iron-rich regions of the lowermost mantle and large low shear velocity provinces (LLSVPs), partitioning in Fe\textsuperscript{2+}-rich compositions has also been investigated \cite{Dorfman2021CompositionHeterogeneities}, showing a strong dependence of $K_D$ on $x^\mathrm{Fe}$.

On the theoretical side, \textit{ab initio} calculations of $K_D$, sometimes combined with semi-empirical modeling, have also been carried out \cite{Muir2016, Xu2017}, including the partitioning between Fp and PPv \cite{morgan2018}. 
The methodologies, compositions, and mineralogy have varied, and differences between predictions remain. 
Several approximations that can impact the final result must be made to this complex thermochemical equilibrium problem between multiple solid solution phases. 
In purely \textit{ab initio} calculations, one can use phonons and the quasiharmonic approximation (PH+QHA) \cite{Wentzcovitch2010a, Zhuang2021} or molecular dynamics and thermodynamic integration (MD+TI) \cite{Frenkel2002UnderstandingSimulation, Freitas2016NonequilibriumLAMMPS} to compute free energies. 
The former does not include intrinsic anharmonic effects, and the latter may not be converged w.r.t simulation cell sizes. 
In the presence of the ISC, control over the iron spin population can be difficult in MD+TI-type calculations. 
The nature of the solid solution modeling is also significant and non-unique \cite{Holmstrom2015SpinDynamics, Sun2022, Mendez2022BroadMantle, Cobden2024Full-waveformMantle}. 
The first studies of the ISC in Fp used ideal solid solution modeling for the HS/LS crossover and Henryian-type modeling for the Mg/Fe solution \cite{Tsuchiya06, Wentzcovitch2009}. 
This is a good approximation for low iron concentrations only, but the very small free energy difference between the HS and LS states in the mixed spin (MS) state enhances the importance of Fe-Fe interactions, demanding a non-ideal HS/LS solid solution modeling level (non-ideal ISC). 
Modeling the magnetic entropy \cite{Tsuchiya06} due to multiple iron spin orientations is also approximate and can dramatically influence the pressure range of the ISC at different temperatures \cite{Cobden2024Full-waveformMantle}. 
Finally, the approximation used in calculating electronic exchange and correlation energy also dramatically impacts the ISC pressure range, especially the Hubbard $U$ \cite{Hsu2010a}. 
Therefore, most calculations choose a reasonable $U$ parameter or adjust it to reproduce 300 K experiments more closely. 
The present study uses a theoretical framework whose approximations have evolved with control and compared with experimental data at every step \cite{Tsuchiya06, Wentzcovitch2009, Hsu, Marcondes2020, Sun2022, Cobden2024Full-waveformMantle}. 
However, approximations for the magnetic entropy, DFT description of the ISC, and Fe/Mg solid solution modeling are important, and their predictive power is not easy to assess at high temperatures where experiments do not exist. 
For instance, the latter forces us to restrict this study to modest iron concentrations in the aggregate, i.e., $x^\mathrm{Fe} < 0.13$. 
More extensive calculations need to be carried out in the future for larger values of $x^\mathrm{Fe}$.

\section{Results and Discussion}
\subsection{Partitioning Coefficient}

\begin{figure}
    \centering
    \includegraphics[width=0.75\linewidth]{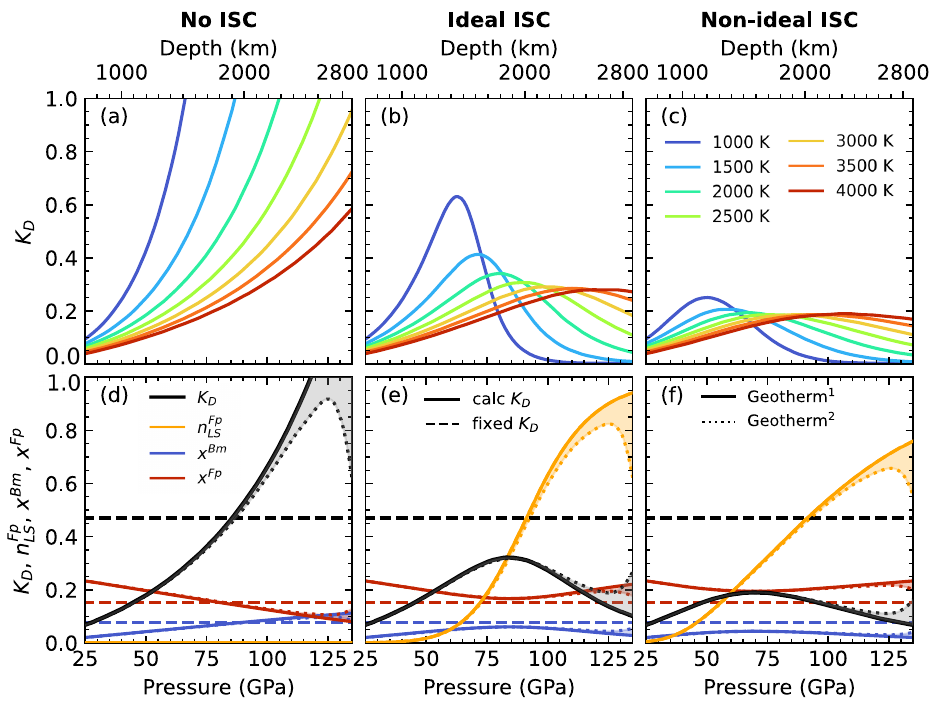}
    \caption{
    $K_D$ predicted by three ISC models: (a, d) no ISC, (b, e) ideal ISC, and (c, f) non-ideal ISC in aggregate pyr\textsuperscript{1} with Bm:Fp volume ratio $4:1$.
    (a-c) show $K_D$ along isotherms; (d-f) show the LS fraction in Fp, $n_\mathrm{LS}^\mathrm{Fp}$, $K_D$, Fe\textsuperscript{2+} concentration in Bm, $x^\mathrm{Bm}$, and in Fp, $x^\mathrm{Fp}$; (d-f) $K_D$ along \textit{geotherm\textsuperscript{1}} \cite{Brown1981} (solid lines) and \textit{geotherm\textsuperscript{2}}\cite{Boehler} (dotted lines).
    Dashed lines in figures (d-f) show $K_D=0.48$ (black), $x^\mathrm{Fp}=0.152$ (red), and $x^\mathrm{Bm}=0.078$ (blue) in the ``fixed $K_D$'' model.
    }
    \label{fig:kd_x-geotherms}
\end{figure}

The detailed methods in our study, including our theoretical framework and \textit{ab initio} calculations, are provided in the supporting materials (see Text ST1).
Using Eq. S1, we predict the iron concentrations of Fp, $x^\mathrm{Fp}$, and Bm, $x^\mathrm{Bm}$, in a pyrolitic aggregate, \textit{pyr\textsuperscript{1}}, with Bm:Fp molar ratio $1:0.6$ (volume ratio 4:1) and average iron concentration $x^\mathrm{Fe} = 0.1$ along isotherms (see Fig. S1 for details).
This is followed by calculations of $K_D$ using Eq. S2 (Figs. \ref{fig:kd_x-geotherms}(a-c)). 
In the absence of the ISC, $K_D$ increases monotonically with pressure but more slowly at higher temperatures (Fig. \ref{fig:kd_x-geotherms}(a)). 
The ISC reverses this trend and produces a maximum in $K_D$ for LS fraction in Fp in the range $0.2<n_\mathrm{LS}^\mathrm{Fp}<0.3$ (Figs. \ref{fig:kd_x-geotherms}(b,c)). 
This $P,T$-dependence of $K_D$ results from the ISC phase diagram, i.e., the rainbow-type diagram \cite{Tsuchiya06}. 
With increasing $T$, the ISC pressure range broadens and shifts to higher $P$, and so does the peak in $K_D$. 
Compared to the ideal ISC, the non-ideal modeling increases the ISC pressure range \cite{Holmstrom2015SpinDynamics, Sun2022, Mendez2022BroadMantle, Cobden2024Full-waveformMantle}, broadens the peak in $K_D$ and lowers its maximum value (see Figs. \ref{fig:kd_x-geotherms}(b,c)). 
Variations of $K_D$ with Fp and Bm proportions and small changes in $x^\mathrm{Fe}$ are not significant and are reported in Fig. S2. 

Figs. \ref{fig:kd_x-geotherms}(d-f) display $n_\mathrm{LS}^\mathrm{Fp}$, $x^\mathrm{Fp}$, $x^\mathrm{Bm}$, and $K_D$ along adiabatic \cite{Brown1981} and non-adiabatic \cite{Boehler} geotherms (\textit{geotherm\textsuperscript{1}} and \textit{geotherm\textsuperscript{2}}, respectively). 
The full pressure and temperature dependence of these quantities are shown in Fig. S3.
Along \textit{geotherm\textsuperscript{1}}, the ideal ISC modeling produces a maximum $K_D\sim 0.32$, maximum $x^\mathrm{Bm}\sim0.06$, and minimum $x^\mathrm{Fp}\sim0.17$, all at $\sim80$ GPa. 
The non-ideal ISC modeling has a maximum $K_D\sim 0.19$, maximum $x^\mathrm{Bm}\sim0.04$, and minimum $x^\mathrm{Fp}\sim0.19$, all at $\sim70$ GPa. 
In the absence of the ISC (Fig. \ref{fig:kd_x-geotherms}(d)) $K_D>1.0$ for $P > 115$ GPa (or $\sim2,650$ km depth) and reaches a maximum value of $\sim0.92$ at $\sim125$ GPa ($\sim2,750$ km depth) along \textit{geotherm\textsuperscript{2}}. 
Results up to 135 GPa are included here for completeness only. 
The PPv transition in Bm will dramatically affect these results. 
In the absence of a model for $K_D$, it is common to assume $K_D\sim0.5$, which results in $x^\mathrm{Fp}\sim0.152$ and $x^\mathrm{Bm}\sim0.078$ (see dashed lines in Figs. \ref{fig:kd_x-geotherms}(d-f) for the ``fixed $K_D$'' model).

\begin{figure}
    \centering
    \includegraphics[width=0.75\linewidth]{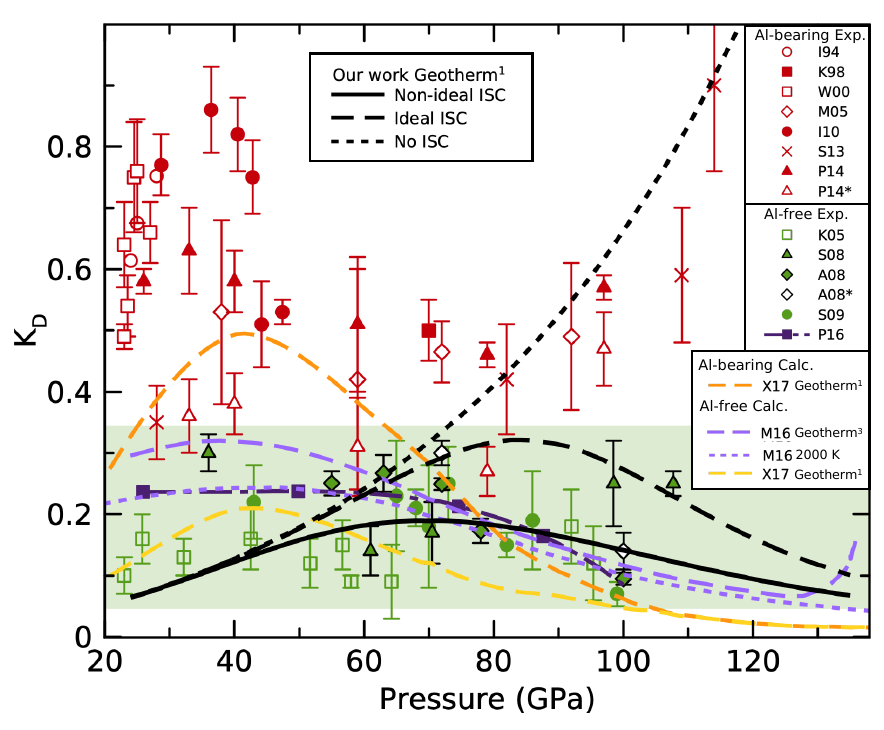}
    \caption{Calculated Fe\textsuperscript{2+} partition coefficient $K_D$ between Bm and Fp compared with experimental (I94 \cite{Irifune2}, K98 \cite{Kesson1998}, W00 \cite{Wood2000}, M05\cite{Murakami2005}, I10 \cite{Irifune2010}, S13 \cite{Sinmyo2013}, P14 \& P14* \cite{Prescher2014}, K05\cite{kobayashi05}, S08 \cite{sinmyo08}, A08 \& A08* \cite{auzende}, S09 \cite{Sakai2009}, P16 \cite{piet}) studies, and computational (M16 \cite{Muir2016}, X17 \cite{Xu2017}) results
    along \textit{geotherm\textsuperscript{1}} \cite{Brown1981} and \textit{geotherm\textsuperscript{3}} \cite{Ono2008ExperimentalMantle}.
    Red and green symbols refer to Al-bearing and Al-free systems, respectively.
    The green-shaded area shows the range of $K_D$ values measured and calculated in Al-free systems.}
    \label{fig:Kd-comparison}
\end{figure}

Fig. \ref{fig:Kd-comparison} compares $K_D$ values predicted by these three ISC models with previous experimental measurements \cite{Irifune2010, Kesson1998, kobayashi05, Sinmyo2013, Prescher2014, Wood2000, auzende, Sakai2009, sinmyo08, Irifune2, Murakami2005, piet} and \textit{ab initio} calculations \cite{Muir2016, Xu2017}.
Measured values vary significantly, even on similar starting materials.
Red markers in Fig. \ref{fig:Kd-comparison} refer to Al-bearing systems. 
At low pressures, $\sim 25$ GPa, garnet breaks down into cCaPv and Al-rich Bm.
Owing to the charge-coupled substitution of Mg\textsuperscript{2+} and Si\textsuperscript{4+} by Fe\textsuperscript{3+} and Al\textsuperscript{3+}, the presence of Al in pyrolite contributes to the iron enrichment in Bm by stabilizing Fe\textsuperscript{3+}. 
This results in a substantially higher $K_D$ than those represented by green markers referring to Al-free systems (mostly San Carlos olivine with $x^\mathrm{Fe}\sim 0.1$).
In Al-bearing pyrolite, $K_D$ seems to peak at P $\sim30$ GPa. 
In Al-free systems, the maximum in $K_D$ seems to decrease from $\sim0.21$ beyond $P > 70$ GPa \cite{piet}. 
Our calculated $K_D$ with the non-ideal ISC model peaks at $\sim 70$ GPa and is consistent with the range of measured values in Al-free systems.

The \textit{ab initio} study by Muir and Brodholt \cite{Muir2016} on a similar Al-free aggregate (Fig. \ref{fig:Kd-comparison}) reported a broader peak in $K_D$  centered at $\sim 40$ GPa along \textit{geotherm\textsuperscript{3}} \cite{Ono2008ExperimentalMantle}, but with a maximum value of $\sim 0.3$, closer to our ideal ISC result. 
The calculation by Xu \textit{et al.} \cite{Xu2017} employed a more approximate thermodynamic framework and predicted a peak in $K_D$ of $\sim 0.6$ for the Al-bearing system and $\sim 0.2$ for the Al-free system at $\sim 43$ GPa along \textit{geotherm\textsuperscript{1}} (Fig. \ref{fig:Kd-comparison}). 
Their maximum value is similar to ours, even though the peak happens 30 GPa lower than ours.
There are too many differences between these studies to assess the origin of these diverse results. 
However, all calculations and measurements of $K_D$ in Al-free systems fall in the same range (green shaded area in Fig. \ref{fig:Kd-comparison}). 

\section{Geophysical Significance}
\subsection{Impact of Iron Partitioning on Seismic Velocities}
\label{section.iron.part.on.v}

The ISC in Fp induces notable elastic anomalies in this phase \cite{Crowhurst,  Antonangeli2011, Marquardt, Wentzcovitch2009, Wu2013} and aggregates containing it \cite{Wu2014}. 
The main anomaly occurs in the bulk modulus, $K$, owing to the anomalous volume contraction throughout the ISC \cite{Wentzcovitch2009}. 
The shear modulus, $G$, is not significantly affected \cite{Marquardt, Wu2013}. The calculated elastic moduli and density of Fp with $x^\mathrm{Fp}=0.1$ agree well with experimental data at 300 K on the same composition (Fig. S4), a good starting point for predicting velocity anomalies in aggregates. 
Using the non-ideal ISC model, the pressure range of the softening anomaly in K broadens, and the anomaly magnitude decreases \cite{Sun2022, Mendez2022BroadMantle, Cobden2024Full-waveformMantle}. 
Here we ask: what is the effect of the Fe\textsuperscript{2+} partitioning on aggregate velocities and density in the presence of the ISC in Fp? 

\begin{figure}
    \centering
    \includegraphics[width=0.75\linewidth]{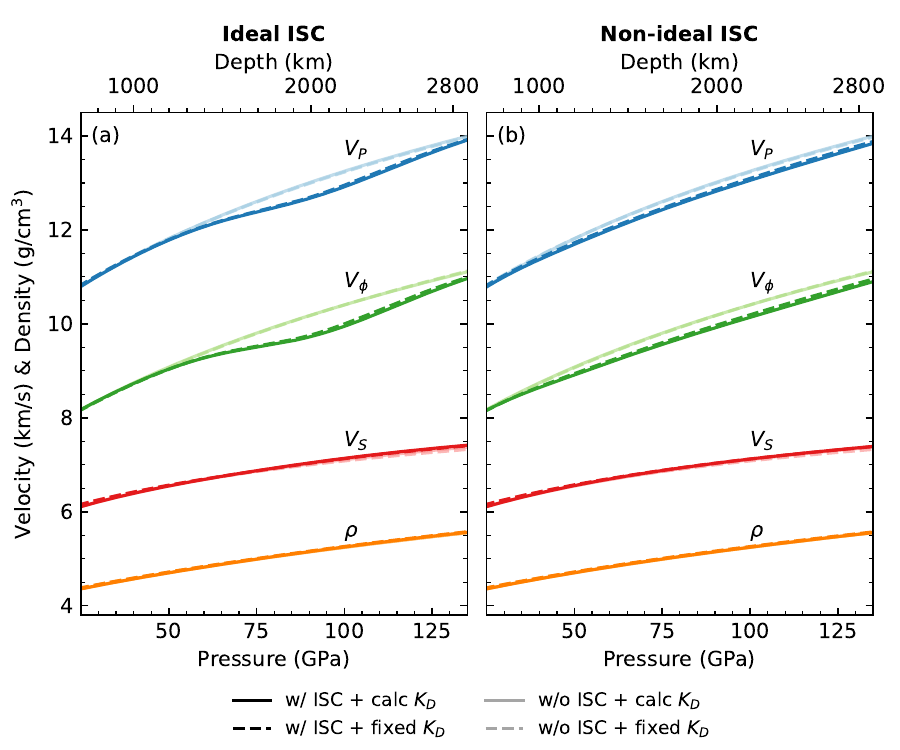}
    \caption{
Calculated P-wave ($V_P$), bulk ($V_\phi$), and S-wave ($V_S$) velocities and density ($\rho$) along \textit{geotherm\textsuperscript{1}} \cite{Brown1981} for a pyrolitic aggregate \textit{pyr\textsuperscript{1}} containing 75 vol \% of Bm, 18 vol \% Fp, 7 vol \% cCaPv, and only Fe\textsuperscript{2+} with $x^\mathrm{Fe} = 0.1$. 
Dark- and light-colored lines indicate aggregates with and without ISC in Fp, respectively.
The ISC effects are modeled as (a) ideal and (b) non-ideal HS/LS solid solutions.
Dashed lines represent aggregates containing Fp undergoing ISC but with fixed $K_D$ (fixed $x^\mathrm{Fp}$ and $x^\mathrm{Bm}$).  
    }
    \label{fig:velocity_variation}
\end{figure}

We compare results on aggregates with variable and fixed $K_D=0.48$ and on aggregates without ISC in Fp (Fig. \ref{fig:velocity_variation}). 
Results predicted by ideal (Fig. \ref{fig:velocity_variation}a) and the non-ideal (Fig. \ref{fig:velocity_variation}b) ISC models lead to the same conclusion: the aggregate in thermochemical equilibrium, with variable $K_D$, $x^\mathrm{Fp}$, and $x^\mathrm{Bm}$ (dark solid lines), has velocities similar to those of an aggregate with these parameters fixed (dark dashed lines). 
Overall, it is the occurrence of the ISC that impacts most strongly aggregate velocities (light vs. dark lines in Fig. \ref{fig:velocity_variation}). 
Therefore, it is essential to model the ISC accurately. 
With the ideal ISC model, the reductions in$ V_\phi$ and $V_P$ along \textit{geotherm\textsuperscript{1}} \cite{Brown1981} are recognizable within the depth range of $\sim1,300 - 2,700$ km ($\sim 60$ GPa$ < P < 125$ GPa).
With the non-ideal ISC model, velocity reductions start at $\sim1,000$ km and persist down to the core-mantle boundary (CMB) ($\sim 40$ GPa $< P < 135$ GPa). 
Both ISC models reduce $V_\phi$ and $V_P$ and slightly increase $V_S$ in the mantle. 
Detailed effects of the ISC and variable $K_D$, $x^\mathrm{Fp}$, and $x^\mathrm{Bm}$, on the velocities and densities of individual phases are shown in Fig. S5. 
We present results up to CMB pressures for completeness, although we do not address the effect of the PPv transition.

\subsection{Impact of Iron Partitioning on Velocity Heterogeneities}
\label{section.iron.part.RSP}

\begin{figure}
    \centering
    \includegraphics[width=1\linewidth]{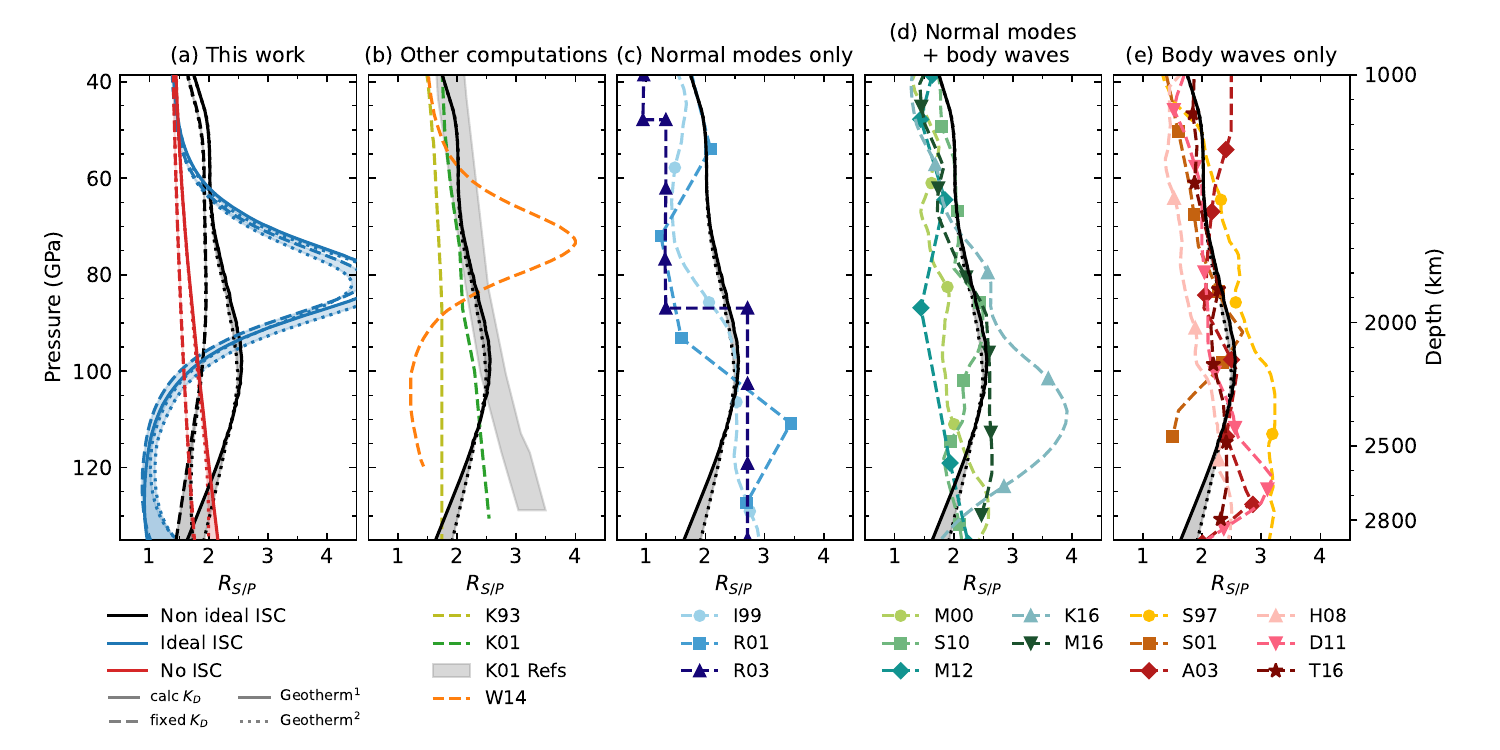}
    \caption{
    Thermally-induced heterogeneity ratio $R_{S/P}$ ($\partial \ln V_S / \partial \ln V_P$), (a) predicted along \textit{geotherm\textsuperscript{1}} and \textit{geotherm\textsuperscript{2}} in \textit{pyr\textsuperscript{1}} using the non-ideal ISC and no ISC models, and
    \textit{pyr\textsuperscript{2}} using the ideal ISC model; 
    (b) comparison with previous estimations based on mineral physics, W93 \cite{Karato1993ImportanceTomography}, mineral physics and seismic tomography, K01 \cite{Karato2001OriginMantle}, and \textit{ab initio} calculations W14 \cite{Wu2014}; 
    values derived from (c) tomographic models based on normal mode data only: 
    I99 \cite{Ishii1999Normal-ModeMantle}, 
    R01 \cite{Romanowicz2001CanMantle}, 
    R03 \cite{Resovsky2003UsingRelationships};
    (d) tomographic models based on body waves + normal modes: 
    M00 \cite{Masters2000TheStructure}, 
    S10 \cite{Simmons2010GyPSuM:Speeds}, 
    M12 \cite{Mosca2012SeismicTomography}, 
    K16 \cite{Koelemeijer2016SP12RTS:Mantle}, 
    M16 \cite{Moulik2016TheMantle};
    (e) tomographic models based on body waves only: 
    S97 \cite{Su1997SimultaneousMantle}, 
    S01 \cite{Saltzer2001ComparingMantle}, 
    A03 \cite{Antolik2003J362D28:Mantle}, 
    H08 \cite{Houser2008ShearWaveforms}, 
    D11 \cite{DellaMora2011LowMantle}, and 
    T16 \cite{Tesoniero2016S-to-PImplications}.
    \textit{Pyr\textsuperscript{1}} aggregate contains
    75 vol \% Bm, 18 vol \% Fp and 7 vol \% cCaPv and $x^\mathrm{Fe} = 0.1$;
    \textit{pyr\textsuperscript{2}} aggregate contains 83 vol \% Bm, 10 vol \% Fp and 7 vol \% cCaPv and $x^\mathrm{Fe} = 0.1$.
    }
    \label{fig:lateral-variations}
\end{figure}

While the effect of the ISC and iron partitioning in radial velocity profiles might be subtle, especially with the non-ideal ISC model, 
it has been pointed out that thermally induced variations in $V_\phi$ and $V_P$, i.e., $R_{\phi/T}$ and $R_{P/T}$, fluctuate significantly. 
The effect was predicted a decade ago \cite{Wu2014} using the ideal ISC model (see Figs. S6(a,b), S7(a,b), and Text ST4). 
Here, we focus on how the ISC solid solution modeling and iron partitioning may affect the $R_{S/P}= \partial \ln V_S / \partial \ln V_P$ ratio, an important parameter in seismology. 
Without iron partitioning (fixed $K_D$ model), the ideal ISC predicts a peak in $R_{S/P}$ centered at $\sim1,800$ km depth (dashed-blue line in Fig. \ref{fig:lateral-variations}(a)). 
Iron partitioning increases the magnitude of this peak from $\sim 4.5$ to $\sim 4.8$ (solid blue line in Fig. \ref{fig:lateral-variations}) owing to the increase in $x^\mathrm{Fp}$ compared to the constant value (see Fig. \ref{fig:kd_x-geotherms}(e)). 
The peak amplitude is only slightly sensitive to the Mg/Si ratio and small variations of the iron content in the aggregate (see Figs. S6 and S7). 

The non-ideal ISC produces a more complex behavior. 
First, its effect starts at $\sim 1,000$ km and persists to CMB depths. 
With fixed $K_D$ (black-dashed line in Fig. \ref{fig:lateral-variations}(a)), $R_{S/P}$ increases slightly and monotonically up to $\sim 1,800$ km depth and then decreases toward CMB depths. 
It varies little between $1.5$ and $2.0$, which reflects the very broad range of the non-ideal ISC throughout the mantle. 
Because it is so broad, the variation in $x^\mathrm{Fp}$ due to iron partitioning (Fig. \ref{fig:kd_x-geotherms}(f)) produces an undulation in an otherwise slowly varying and featureless depth-dependent $R_{S/P}$ (solid-black vs. dashed-black lines in Fig. \ref{fig:lateral-variations}(a)). 
The undulation depends on the Mg/Si ratio and iron content, $x^\mathrm{Fe}$ (see the variation in $R_{P/S} = 1/R_{S/P}$ with Mg/Si ratio in Fig. S7(e)).

Except for this undulation, our predicted $R_{S/P}$ in a pyrolitic aggregate agrees well down to $\sim 2,400$ km with values Karato and Karki \cite{Karato2001OriginMantle} extracted from tomography models \cite{Masters2000TheStructure, Su1997SimultaneousMantle, Robertson1996ConstraintsPhysics} (grey band in Fig. \ref{fig:lateral-variations}(b)).
Without the ISC (red curves in Fig. \ref{fig:lateral-variations}(a)), our predicted range of $R_{S/P}$ (with or without partitioning) is similar to the values they predicted for MgO and MgSiO\textsubscript{3}. 
These authors added anelastic effects to bring mineral physics results in better agreement with $R_{S/P}$) derived from tomography (dashed-green and broad grey band in Fig. \ref{fig:lateral-variations}(b)). 
The ISC-induced bulk modulus softening and iron partitioning effects can bring the predicted $R_{S/P}$ in agreement with tomography values, at least down to $2,400$ km.
Anelastic effects are not needed. In fact, the ability of the ISC to better explain mantle velocities in the $1,800 - 2,500$ km depth range \cite{Cobden2024Full-waveformMantle} suggests anelastic effects are counterproductive in the description of mantle velocities. 
The increase of our $R_{S/P}$ with depth is also in good agreement with the behavior extracted from several tomography models (see Figs. \ref{fig:lateral-variations}(c-e)), reporting undulating increases with depth. 
Particularly intriguing is the similarity between the undulation in our $R_{S/P}$ (black-solid line in Fig. \ref{fig:lateral-variations}) and those inferred from models I99 \cite{Ishii1999Normal-ModeMantle}, R01 \cite{Romanowicz2001CanMantle}, S10 \cite{Simmons2010GyPSuM:Speeds}, and M16 \cite{Moulik2016TheMantle}. 
This behavior of $R_{S/P}$ is a consequence of the iron partitioning caused by the ISC. 
See Fig. S8 for a more detailed comparison with tomography models.

\subsection{Consequences for Plume and Slab Images}
\label{section.fe.part.plume.slab}
\begin{figure}
    \centering

    \includegraphics[width=1\linewidth]{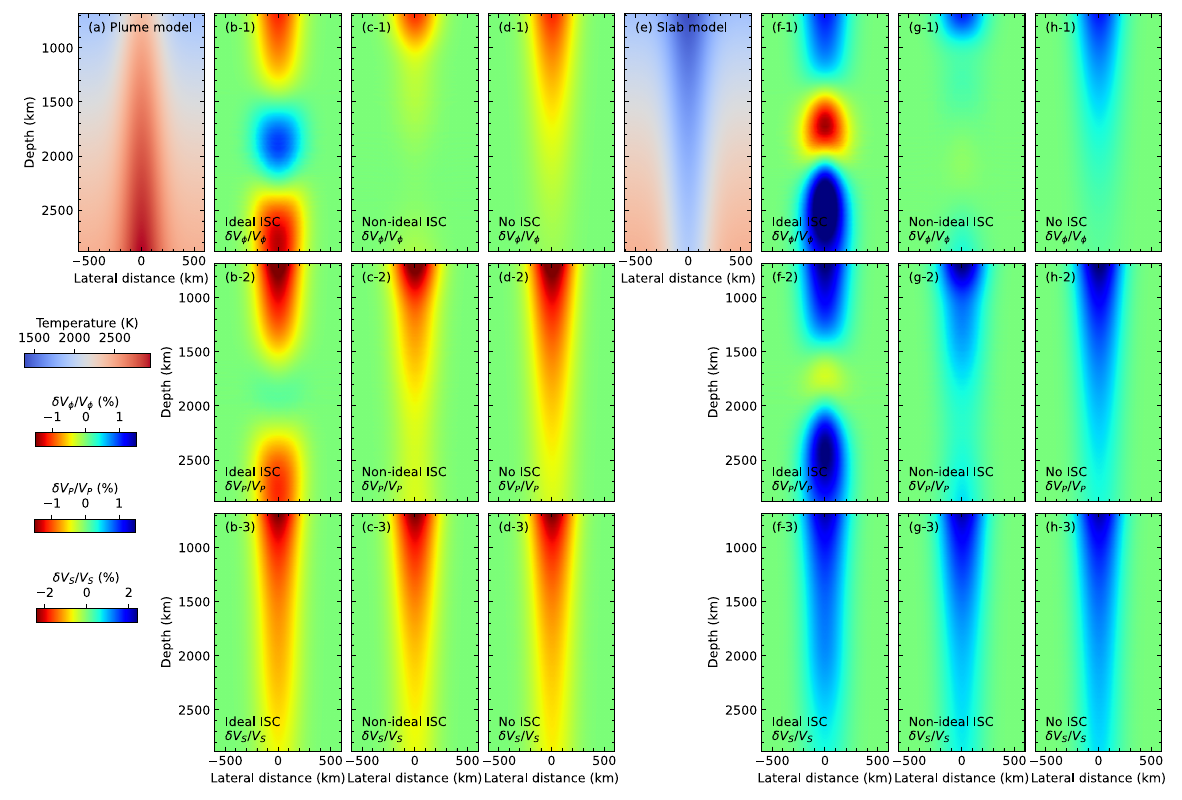}
    \caption{Calculated bulk $\delta V_\phi/V_\phi$ (\%),  P-wave $\delta V_P/V_P$ (\%), and S-wave velocity heterogeneities $\delta V_S/V_S$ (\%) induced by lateral temperature variations using ideal, non-ideal, and no ISC modeling including iron partitioning effects in \textit{pyr\textsuperscript{1}} aggregate. 
    (a) Plume and (e) slab temperatures are 500 K above and below the \textit{ambient mantle geotherm\textsuperscript{1}} \cite{Brown1981}.
    Lateral temperature variations are vertically uniform and have Gaussian distributions with standard variance $\sigma=150$ km for reference. 
    \textit{Pyr\textsuperscript{1}} contains 75 vol \% Fe\textsuperscript{2+}-Bm, 18 vol \% Fp and 7 vol \% cCaPv.
    Similar images with fixed $K_D$ or along \textit{geotherm\textsuperscript{2}} \cite{Boehler} are shown in Figs. S10 and S12, respectively.
    }
    \label{fig:plume-slab}
\end{figure}

To understand the effect of the ISC and iron partitioning on the appearance of slabs and plumes in tomography maps, we sketch velocities produced by vertically uniform colder and warmer thermal structures with temperature distributions $T=T_{\mathrm{geotherm}}(d) + \Delta T g(x)$ (Fig. \ref{fig:plume-slab}) in an isochemical pyrolitic mantle. 
$T_{\mathrm{geotherm}}$ is the \textit{adiabatic mantle geotherm\textsuperscript{1}} \cite{Brown1981}, $d$ is depth, $\Delta T$ is the temperature variation towards the center of the plume ($+ 500$ K) or slab ($-500$ K) features, and $g(x)$ is a Gaussian distribution with the standard variance $\sigma=150$km for reference (see the central $T-P$ profiles in Fig. S9). 
Fig. \ref{fig:plume-slab} shows these images including iron partitioning effects. 
Corresponding images with no iron partitioning, or, along a \textit{non-adiabatic geotherm\textsuperscript{2}} \cite{Boehler} are shown in Figs. S10 and S12, respectively.  

First, in the absence of the ISC in Fp (no ISC model), the appearance of plumes and slabs represented as $\Delta V \%$ are qualitatively similar in $S$-, $P$-, or $\phi$- models (Figs. \ref{fig:plume-slab}(d-1,2,3; h-1,2,3)). 
This is because $R_{X/T}^{\mathrm{Pyr}}=\partial \ln V_X^\mathrm{Pyr}/\partial T$, for $X = S$, $P$, or $\phi$, have similar depth dependence (see Figs. S6(g,h)). 
Their ratios vary little and monotonically throughout the mantle (Figs. S7(g,h)). 
Iron partitioning has a minor qualitative impact on these images (compare Fig. \ref{fig:plume-slab} and Fig. S10 with fixed $K_D$).

Second, in the ideal ISC model, $\phi$- and $P$-images are severely disrupted in the $1,300 - 2,400$ km depth range in the slab image (Figs. \ref{fig:plume-slab}(f-1,2)). 
The plume image is disrupted 300 km deeper (Figs. \ref{fig:plume-slab}(b-1,2)), as expected \cite{Wu2014}. 
In $\phi$-images, $\Delta V_\phi$ can even reverse (Figs. \ref{fig:plume-slab}(b-1; f-1)) with $R_{\phi/T}$ becoming positive in this depth range (see Figs. S6(a,b)). 
In the absence of iron partitioning, such disruptions are less accentuated (see Figs. S10(b-1,2; f-1,2)). 
S-images are almost unaffected by the ISC (compare Figs. \ref{fig:plume-slab}(f-3; h-3) for slab, (b-3; d-3) for plume) or by iron partition (compare Figs. \ref{fig:plume-slab}(b-3; h-3) with Fig. S10(b-3; h-3)).

Third, in the non-ideal ISC model, these strong velocity disruptions change into $\phi$-, $P$-, and $S$-images fading more similarly with depth for plume and slab (Figs. \ref{fig:plume-slab}(c-1,2,3; g-1,2,3)). 
Fading depths increase from $\phi$- to $P$- to $S$-images, with plumes fading at greater depths than slabs.
Without iron partitioning effects, these velocity fadings are less accentuated at all depths (compare these images with corresponding ones in Fig. S10). 
Although mild, the iron partitioning effect is noticeable on thermally induced velocity heterogeneities.

Except for the non-realistic ideal ISC model, all $\phi$-, $P$-, or $S$-images produced by vertically uniform thermal structures fade with increasing depth, even in models with no ISC and no iron partitioning (check Figs. S10(d-1,2,3; h-1,2,3)). 
This general effect results from a similar decrease in magnitude in $R_{\phi/T}$,  $R_{P/T}$, and $R_{S/T}$ toward the deep mantle (see Figs. S6(d-i)), irrespective of the geotherm (see Fig. S12 for similar images along the \textit{non-adiabatic geotherm\textsuperscript{2}} \cite{Boehler}).
Deviations from such behavior strongly suggest other factors not included in this analysis, e.g., chemical/mineralogical heterogeneities or stratification, PPv transition, etc.

\section{Summary and Conclusions}
We have presented a systematic \textit{ab initio} study of the ferrous iron (Fe\textsuperscript{2+}) partitioning coefficient, $K_D$, in a typical pyrolitic aggregate with Fp undergoing iron spin crossover (ISC) under pressure. 
We have calculated the effect of the ISC on $K_D$ and of the latter on aggregate velocities and the thermally induced heterogeneity ratios $R_{S/P}$($\partial \ln V_S/\partial \ln V_P$) at mantle conditions. 
We have used three ISC models: ideal and non-ideal HS/LS solid solution models and no ISC at all. 
The Mg/Fe solid solution problem was treated at the Henryian solution level in all three cases.
This exercise was done to clarify whether the previously predicted strong effect of the ISC on mantle velocities and lateral heterogeneities \cite{Wu2014} persists with the non-ideal ISC model and whether it is affected by iron partitioning.

First, the ISC effect on aggregate velocities dominates over the iron partitioning effect.
The latter is considerably more subtle, irrespective of how the HS/LS solid solution problem is addressed. 

Second, both ideal and non-ideal ISC models predict $K_D$ within the range of measured values for Al-free aggregates within uncertainties. 
Other calculations using different methods \cite{Muir2016, Xu2017} also predicted $K_D$’s within the same range, but their predicted $K_D$ peak at $\sim 30 - 40$ GPa is at a pressure well below our peak pressure, $70$ GPa. 
Our peak pressure correlates well with the pressure beyond which $K_D$ starts to decrease in measurements by Piet \textit{et al.} \cite{piet}. 

Third, there are qualitative changes in the non-ideal ISC effect on mantle velocities and heterogeneities compared to the previous ideal ISC predictions \cite{Wu2014}. 
The former has a very broad effect on $V_\phi$ and $V_P$ in the mantle, making the ISC effect look like a simple reduction in these velocities from 1,000 km to CMB depth.
It also removes the peak in the important seismological parameter $R_{S/P}$ predicted with the ideal ISC model.
This was a puzzling discrepancy between \textit{ab initio} predictions of $R_{S/P}$ and values extracted from tomography.
Despite the success of the non-ideal ISC model, it must be pointed out that the ideal ISC effect seems to manifest qualitatively correctly in some regions in the mantle \cite{Shephard2021SeismologicalMantle, Trautner2023CompressibilityTomography, Cobden2024Full-waveformMantle}.
Particularly, the depth ranges of predicted disruptions in $V_P$ structures correlate well with plume images under hot spots in some P-models \cite{Nolet2006PlumeTomography, Zhao2007SeismicPlumes, Boschi2007}. 
The fading of plate images in P-models in a similar depth range is also well documented \cite{vanderHilst1999CompositionalModel, Shephard2021SeismologicalMantle}. 
These correlations between ideal ISC predictions and some seismological observations might not be accidental. 

Finally, the non-ideal ISC model predicts $R_{S/P}$ in much better agreement with values inferred from tomography. 
The Fe\textsuperscript{2+} partitioning effect in a homogeneous pyrolitic aggregate is a non-monotonic (undulated) increase of $R_{S/P}$ with depth along the mantle geotherm. 
All tomography models cited in Fig. \ref{fig:lateral-variations} produce undulated increases in $R_{S/P}$ with depth until $\sim 2,300$ km. 
Many factors not considered here, e.g., chemical heterogeneities or stratification, PPv transition, ISC in Fe\textsuperscript{3+}, etc., as well as possible upgrades in the \textit{ab initio} molding of the ISC, should change the behavior of $R_{S/P}$. 
Nevertheless, similar undulations in $R_{S/P}$ inferred from some tomography models (I99 \cite{Ishii1999Normal-ModeMantle}, R01 \cite{Romanowicz2001CanMantle}, S10 \cite{Simmons2010GyPSuM:Speeds}, M16 \cite{Moulik2016TheMantle} in Figs. \ref{fig:lateral-variations} and S8) are particularly intriguing.

%
%



\section{Open Research}

The authors comply with the AGU’s data policy, and the data sets of this paper are available at \url{https://doi.org/10.6084/m9.figshare.25262872}.





\acknowledgments

This work was supported in part by the National Science Foundation award EAR-2000850. This work used SDSC Dell Cluster with AMD Rome HDR IB at Expanse through allocation DMR180081 from the Advanced Cyberinfrastructure Coordination Ecosystem: Services \& Support (ACCESS) program, which is supported by National Science Foundation grants \#2138259, \#2138286, \#2138307, \#2137603, and \#2138296.


%
%



\clearpage
\bibliography{references}

%
%
%
%
%

\end{document}